\documentclass[a4paper,11pt]{article}
\usepackage{pos}

\usepackage{lineno}

\title{Performance of the SoLid electron anti-neutrino detector}

\author*[a]{Noë Roy}
\author{on behalf of the SoLid collaboration.}
\affiliation[a]{Université Paris-Saclay, CNRS/IN2P3, IJCLab, 91405 Orsay, France}
\emailAdd{roy@lal.in2p3.fr}

\abstract{The SoLid collaboration operates since 2018 a 1.6 ton neutrino detector at the Belgian BR2 reactor, with as main goal the search for the oscillation of electron anti-neutrinos to a sterile state. The highly segmented SoLid detector employs a novel hybrid scintillation technology based on PVT scintillator in combination with a $^6$LiF:ZnS(Ag) screens containing $^{6}\mathrm{Li}$ isotopes. The experiment has demonstrated a channel-to-channel response that can be controlled to the level of a few percent, an energy resolution of better than 14$\%$ at 1 MeV, and a determination of the interaction vertex with a precision of 5 cm. This contribution discusses the technology choices that were made for SoLid experiment and the calibration performances required. The ongoing upgrade program of the detector and the expected associated improvements in performance will also be presented.}

\FullConference{%
  40th International Conference on High Energy physics - ICHEP2020\\
  July 28 - August 6, 2020\\
  Prague, Czech Republic (virtual meeting)
}

\begin{document}
\maketitle


\section{Physics motivation}\label{sec:motivation}

The SoLid experiment was designed to study two main anomalies. The first one comes from the so called reactor anti-neutrino anomaly that shows a 3$\sigma$ deficit of anti-neutrino rate at very short baseline from nuclear reactors \cite{Gariazzo_2017}. To this anomaly, $^{235}\mathrm{U}$ may be the primary contributor \cite{235U}. A 3$\sigma$ neutrino deficit was also found during the calibration of the GALLEX and SAGE solar neutrino experiments \cite{gallium}. The existence of a light sterile neutrino state with a mass around 1 eV could account for these observations. The second anomaly comes from Double Chooz, Daya Bay and Reno experiments. Those experiments all showed a distortion of the reactor antineutrino energy spectrum around 5 MeV \cite{double_chooz, daya_bay, reno}. This observation could be due to non linearities of the energy response of the detectors \cite{Mention_2017} or incomplete knowledge of anti-neutrino production in reactor core.

As a very short baseline reactor anti-neutrino experiment and with a linear energy response, the SoLid experiment was designed to study those anomalies. The detector is located at the BR2 research reactor at a distance between six and nine meters of a highly enriched $^{235}\mathrm{U}$ fuel core. An oscillation analysis in the energy-length phase space will be performed to probe the sterile neutrino hypothesis.

\section{SoLid detector}\label{sec:detector}
The interaction of interest in SoLid is the inverse beta decay (IBD):

\begin{equation}
\bar{\nu_e} + p \rightarrow e^{+} + n
\end{equation}
The signal of interest is composed of a spatial and time coincidence between a prompt energy deposit from the positron and a delayed signal after the neutron thermalisation and capture. Due to the proximity of the reactor core and the low overburden in reactor experiments, important atmospheric and reactor induced backgrounds are faced. For signal determination and background reduction, a novel technology has been developed using a highly segmented plastic scintillator (PVT) detector. It allows the IBD selection through the exploitation of the signal topologies. It is also used to reduce the background that can mimic the positron-neutron coincidences. The PVT was also chosen for its linear energy response. The total detector is composed of 50 detection planes, each plane being composed of 16$\times$16 detection cells.
This segmentation comes with a price, 12800 detection cells have to be calibrated to achieve the uniformity required for an oscillation analysis.
The detection cell in SoLid is composed of a $5\times5\times5~ \mathrm{cm}^3$  organic PVT cube as a neutrino target for the positron detection. Two faces of a cube are covered with inorganic $^6$LiF:ZnS(Ag) screens for neutron detection after thermalisation:
\begin{equation}
n+{}^{6}\mathrm{Li} \rightarrow \alpha + {}^{3}\mathrm{H}
\end{equation}
The neutron capture by the $^{6}\mathrm{Li}$ will result in the emission of an alpha particle that will induce the ZnS scintillation. The cubes are optically isolated with Tyvek layers and the signal is readout by four wavelength shifting fibres connected to a Multi-Pixel Photon Counter (MPPC) and a mirror at each end. The MPPC will count the number of electronic avalanches caused by the scintillation (called pixel avalanche PA). The two scintillation signals have different shapes, the ZnS signal (NS) is composed of a long succession of peaks on a few $\mathrm{\mu s}$ while the PVT signal (ES) is a sharp peak. This feature is used for a pulse shape discrimination for the signal of interest \cite{Abreu_2017}.

\begin{figure}
	\centering
	\includegraphics[width=.7\textwidth]{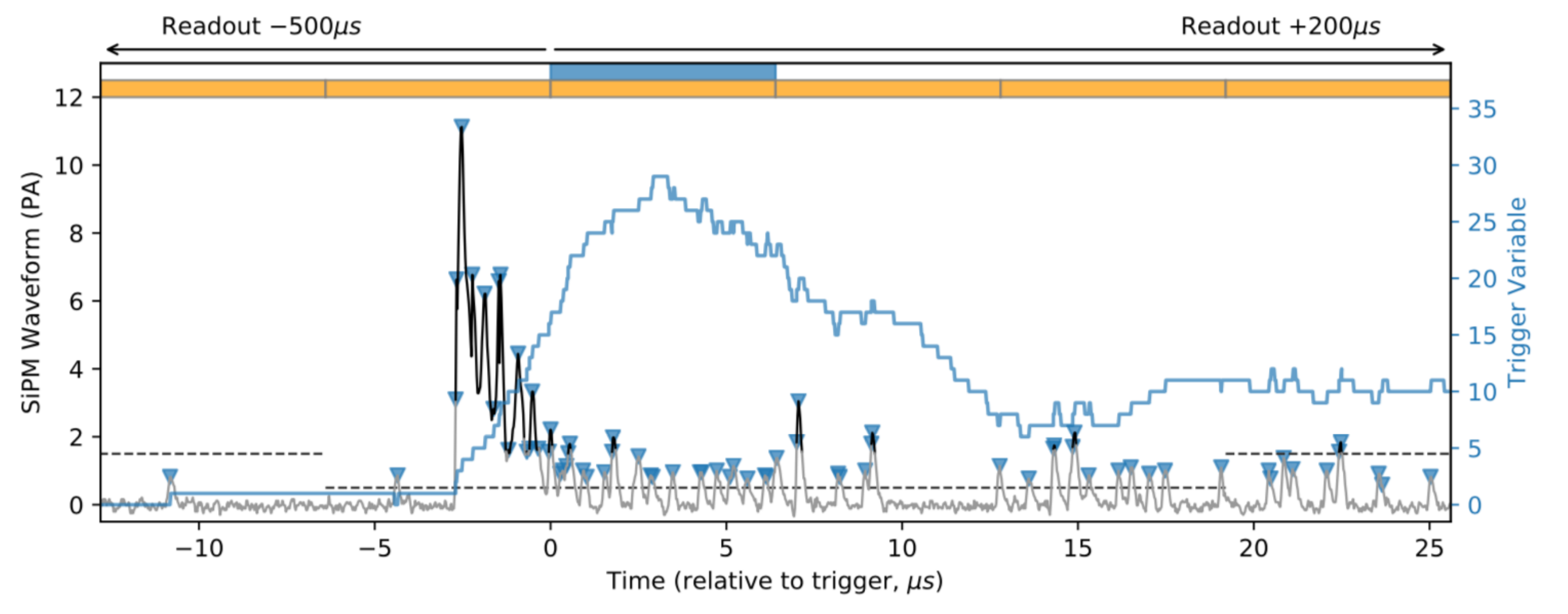}
	\caption{\small Example of neutron waveform (black) \cite{readout}. The dashed lines show the zero suppression threshold. The value of the neutron trigger variable (i.e number of peaks in the rolling time window), using algorithm parameters optimised for physics mode, is shown in blue. The rectangles at the top of the figure denote the blocking of waveform samples, with the triggered block highlighted in blue.}
	\label{fig:trigger}
\end{figure}

Due to the important backgrounds, a dedicated neutron trigger has been set based on a peak counting algorithm, see Fig.~\ref{fig:trigger}, for a more efficient selection of the events of interest around a neutron signal. This trigger allowed the readout of $\pm$ 3 planes around the neutron signal, in a time buffer of [-500 $\mu\mathrm{s}$, 200 $\mu\mathrm{s}$] and a low threshold of 1.5 PA, around 60 keV.

\section{Detector calibration}\label{sec:calibration}
An automated system was constructed to achieve the 12800 detection cells calibration. It was designed to put radioactive sources inside the detector in gaps between planes momentary created for calibration purpose. Two types of calibration have been performed. One for neutron reconstruction efficiency with a neutron source and one for energy calibration with gamma sources.
\subsection{Neutron calibration}
\begin{figure}
 	\centering
 	\includegraphics[width=.65\textwidth]{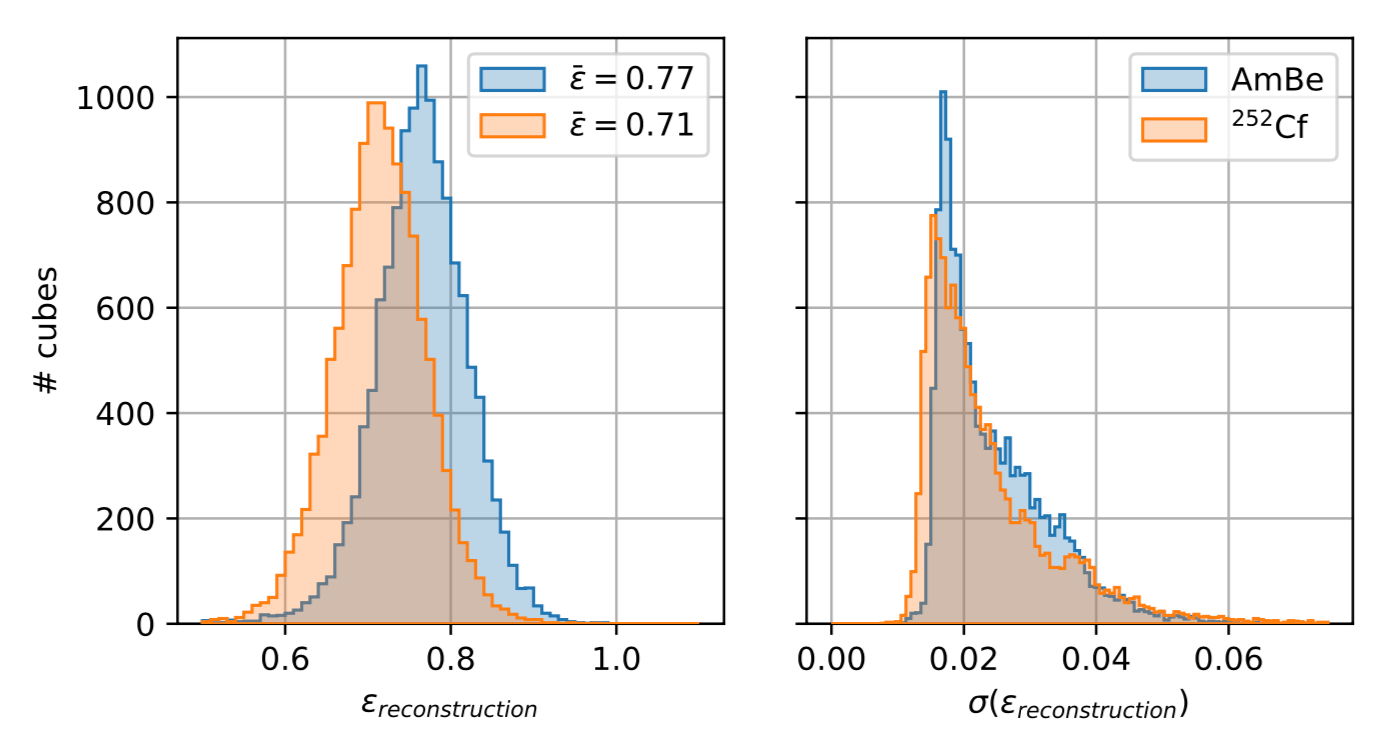}
 	\caption{\small (Left) Neutron reconstruction efficiency for the 12800 cells. (Right) Statistical error on the neutron efficiency measurement for the 12800 cells \cite{detector_paper}.}
 	\label{fig:ns_eff}{\tiny }
 \end{figure}

The neutron calibration was done using an AmBe and a $^{252}\mathrm{Cf}$ source. This calibration confirmed the good ES/NS discrimination with a NS trigger purity above \mbox{99$\%$}. The detection efficiency was measured per cube with an average of 73.9 [+4.0 -3.3]$\%$, as shown on the Fig.~\ref{fig:ns_eff}.

\subsection{Energy calibration}
\begin{figure}
	\centering
	\includegraphics[width=.42\textwidth]{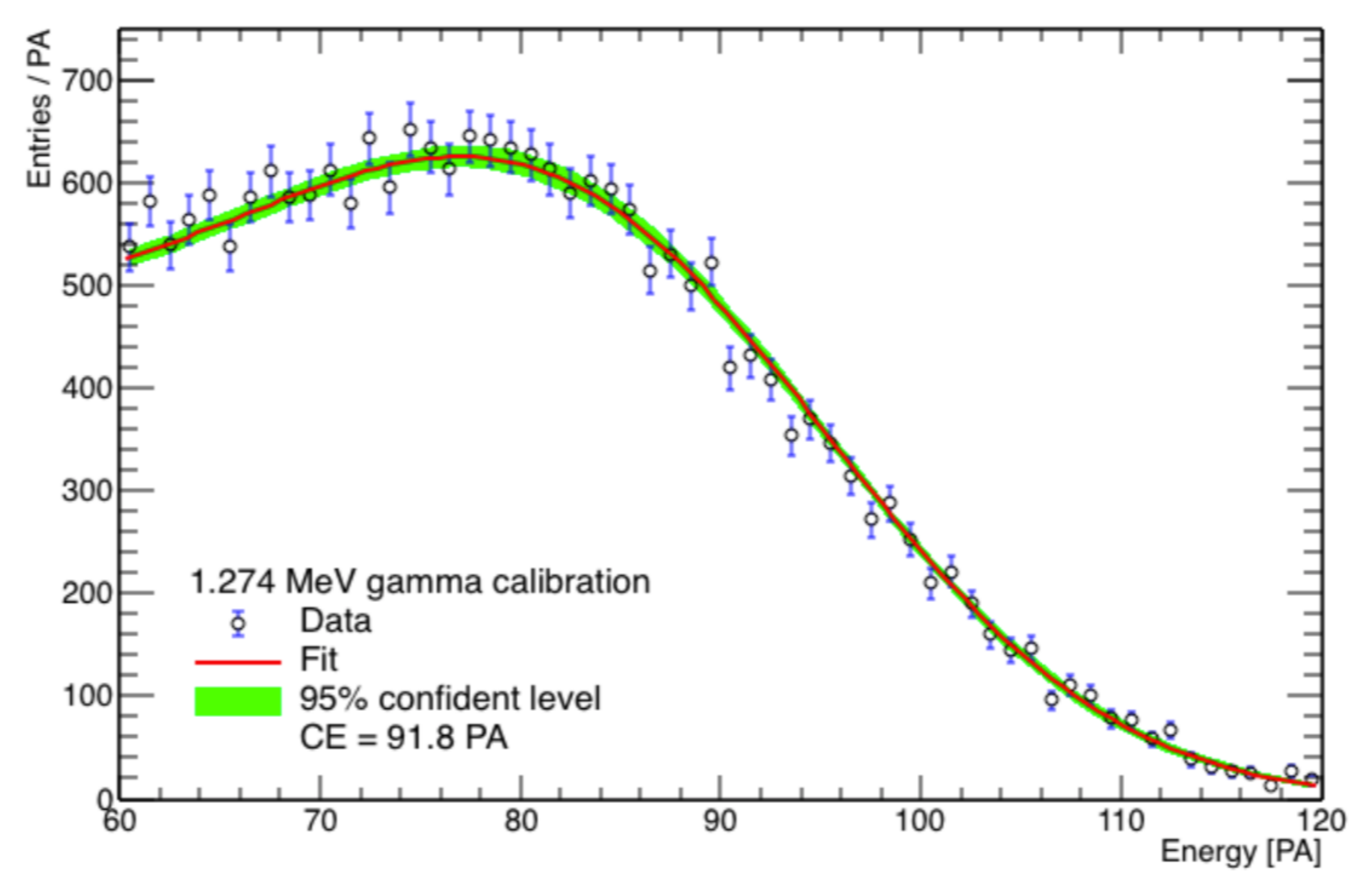}
	\includegraphics[width=.47\textwidth]{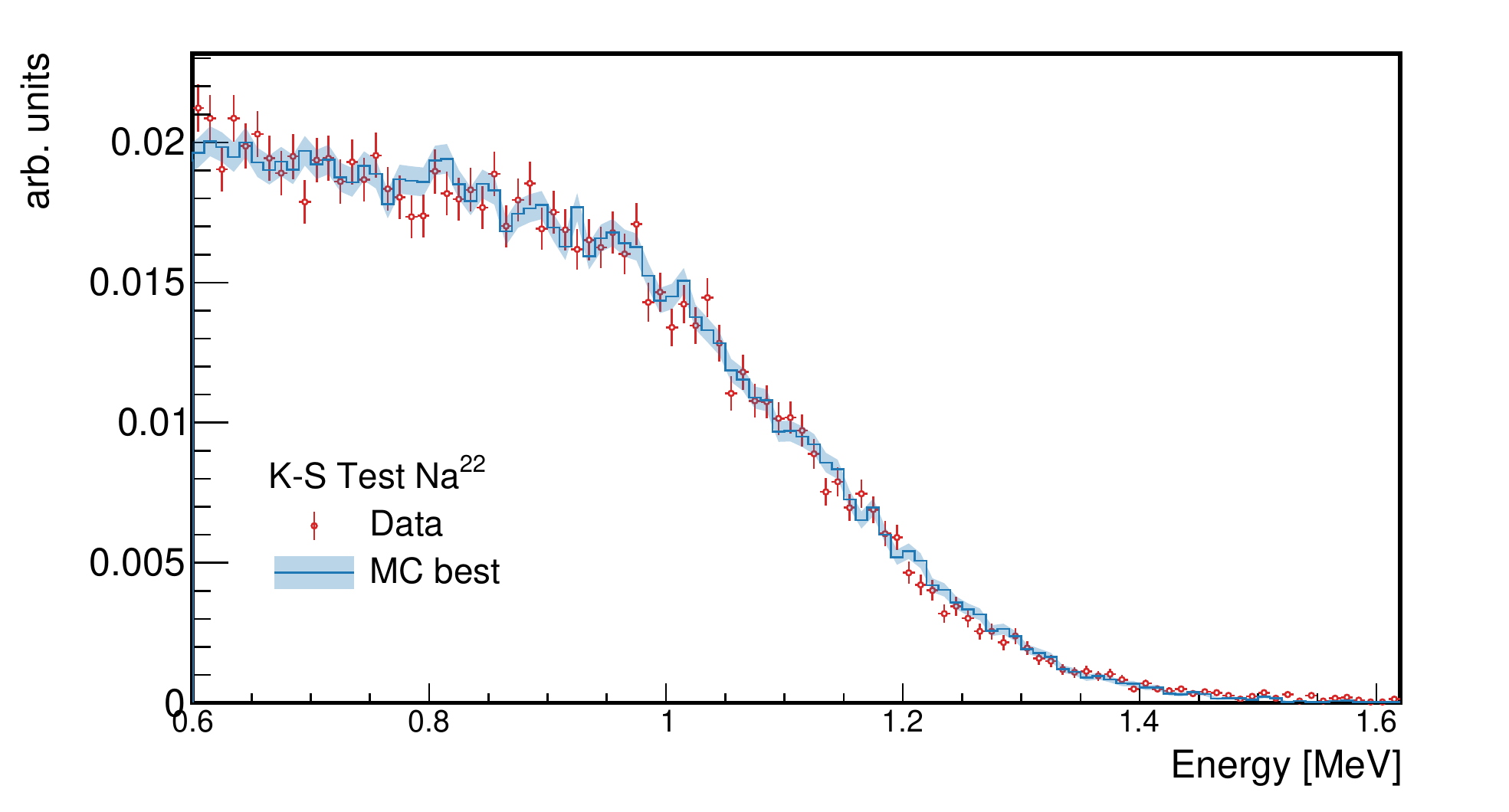}
	\caption{\small Example of light yield measurement obtained with $^{22}\mathrm{Na}$ source by using either a fit of the Compton edge profile (Left) or a Kolmogorov-Smirnov test between Geant4 true energy deposits convoluted with a given energy resolution and data (Right).}
	\label{fig:CE}
\end{figure}

The energy calibration was performed using $^{22}\mathrm{Na}$, $^{207}\mathrm{Bi}$ and AmBe gamma sources. Due to the small size of the detection cells, the gammas do not deposit all their energies in the cubes and the energy calibration has to be performed with Compton edges (CE). Two methods have been developed to determine the light yield (number of PA measured per MeV deposit) and the energy resolution. The first one is an analytical fit based on the Klein-Nishina cross section \cite{Kudomi:1999px}. The second one is based on a Kolmogorov-Smirnov test to compare Data and true Geant4 numerically convoluted energy distributions, see Fig.~\ref{fig:CE}. Light yield of all cubes has been calibrated with a \mbox{3$\%$} precision due to discrepancies between the two methods and reconstruction inefficiencies at low energy.

\begin{figure}
	\centering
	\includegraphics[width=.49\textwidth]{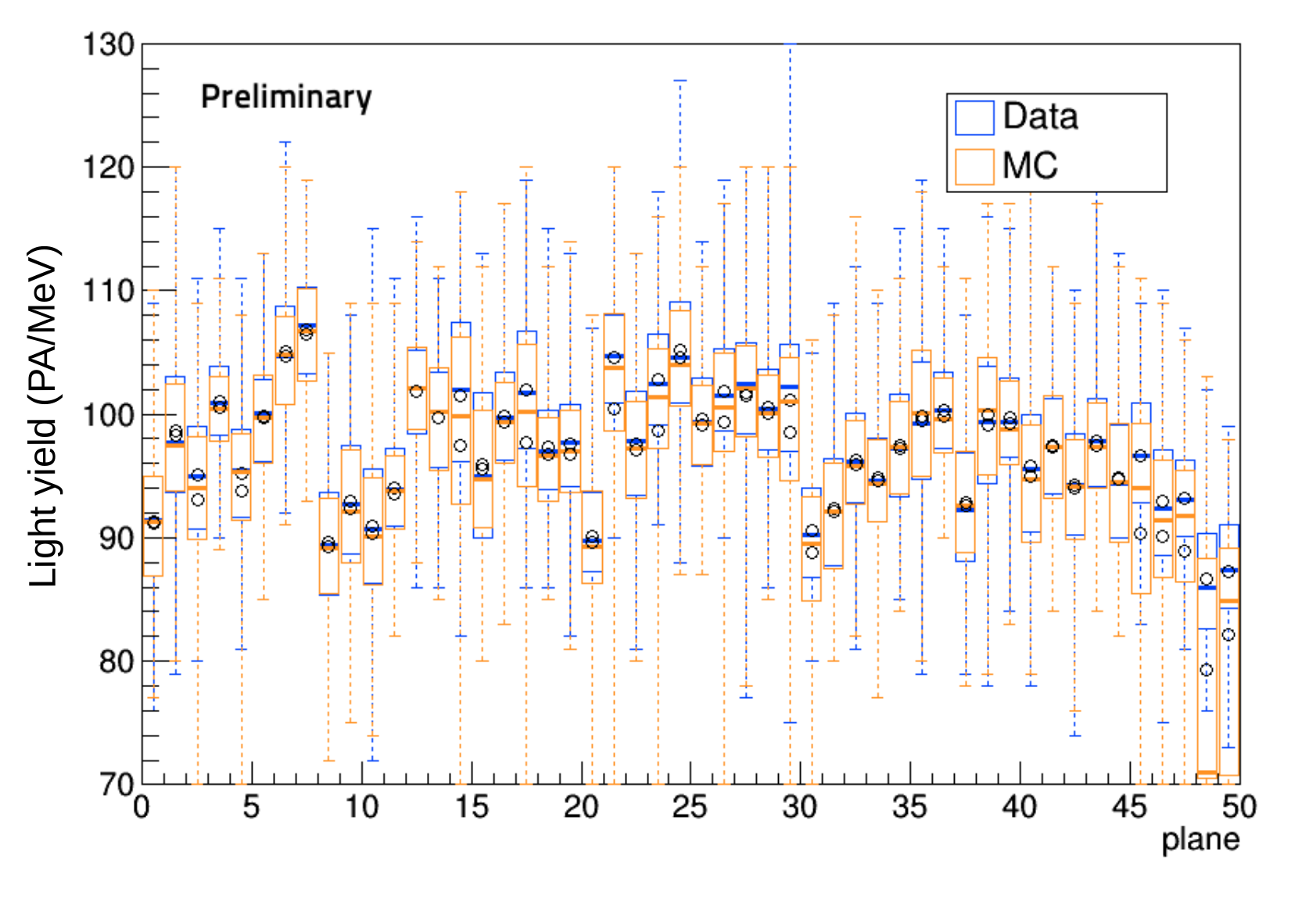}
	\includegraphics[width=.49\textwidth]{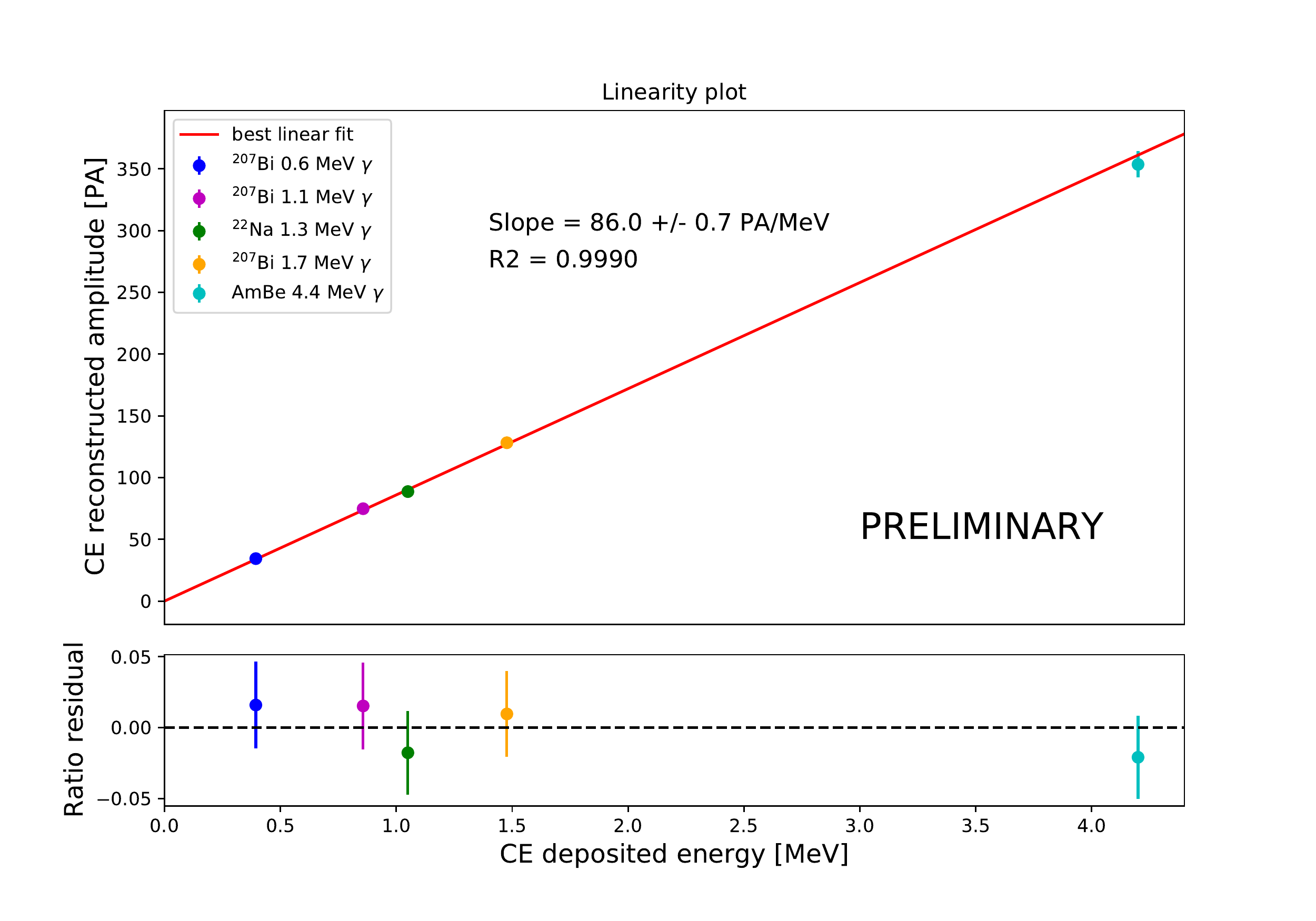}
	\caption{\small Light yield measured in the 50 planes of the SoLid detector (Left). Light yield linearity tested for a subset of 146 cubes (Right). The averaged value is lower than the averaged light yield in the detector due to the subset of cubes tested.}
	\label{fig:LY}
\end{figure}

To ensure the homogeneity of the detector, all the MPPC responses have been equalized at the \mbox{1$\%$} level, individual fibre attenuations have been measured and the optical coupling between the MPPCs and the fibres have been quantified and corrected. After the corrections, we observe a light yield variation of \mbox{3$\%$} within a plane. All these variations have also been reproduced in simulation, as shown on the Fig.~\ref{fig:LY} (Left). The averaged light yield measured with the $^{22}\mathrm{Na}$ in the detector was 96 PA/MeV (this figure does not take into account a \mbox{20$\%$} cross talk of the MPPCs \cite{readout}) for a stochastic energy resolution of \mbox{12$\%$} at 1 MeV. The linearity of the energy response of the detector have been verified for a subset of cubes in the detector and is within a few percent in the [0.5 - 4] MeV region, see Fig.~\ref{fig:LY} (Right). 
Those levels of homogeneity and linearity in the energy response of the detector are key points for the sterile neutrino analysis and $^{235}\mathrm{U}$ induced antineutrino energy measurement.

\subsection{Calibration validation}

\begin{figure}
 	\centering
 	\includegraphics[trim= 0 5 0 0, width=.47\textwidth]{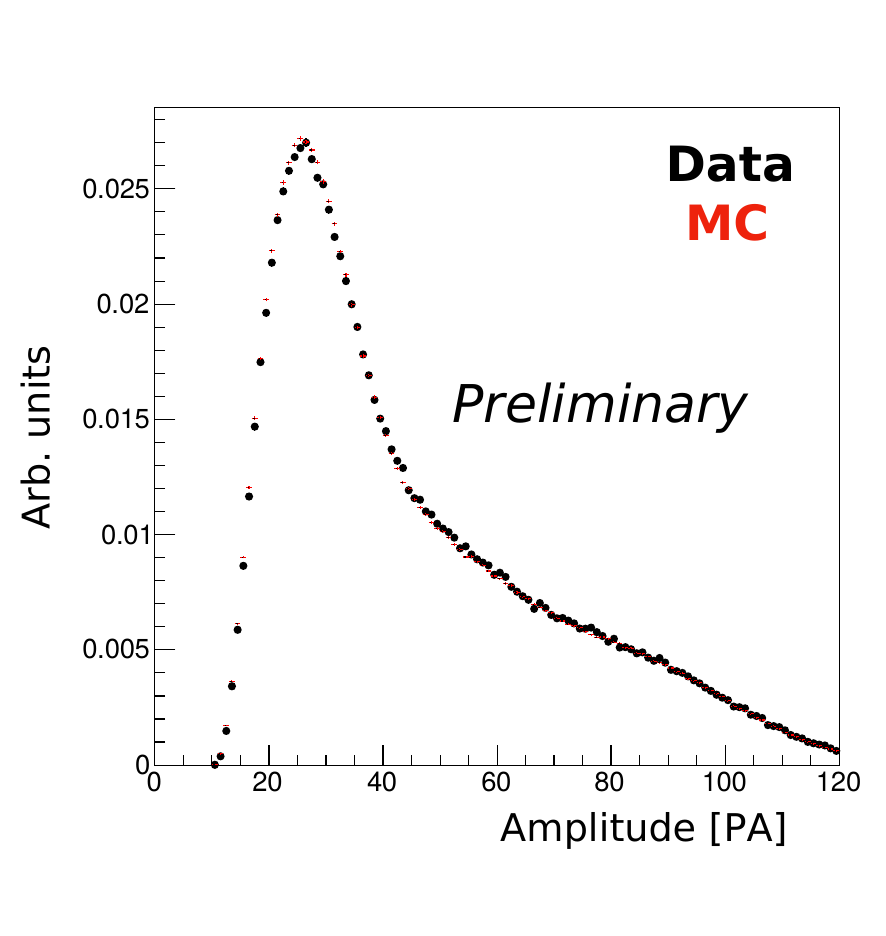}
 	\includegraphics[trim= 0 0 0 2000, width=.49\textwidth]{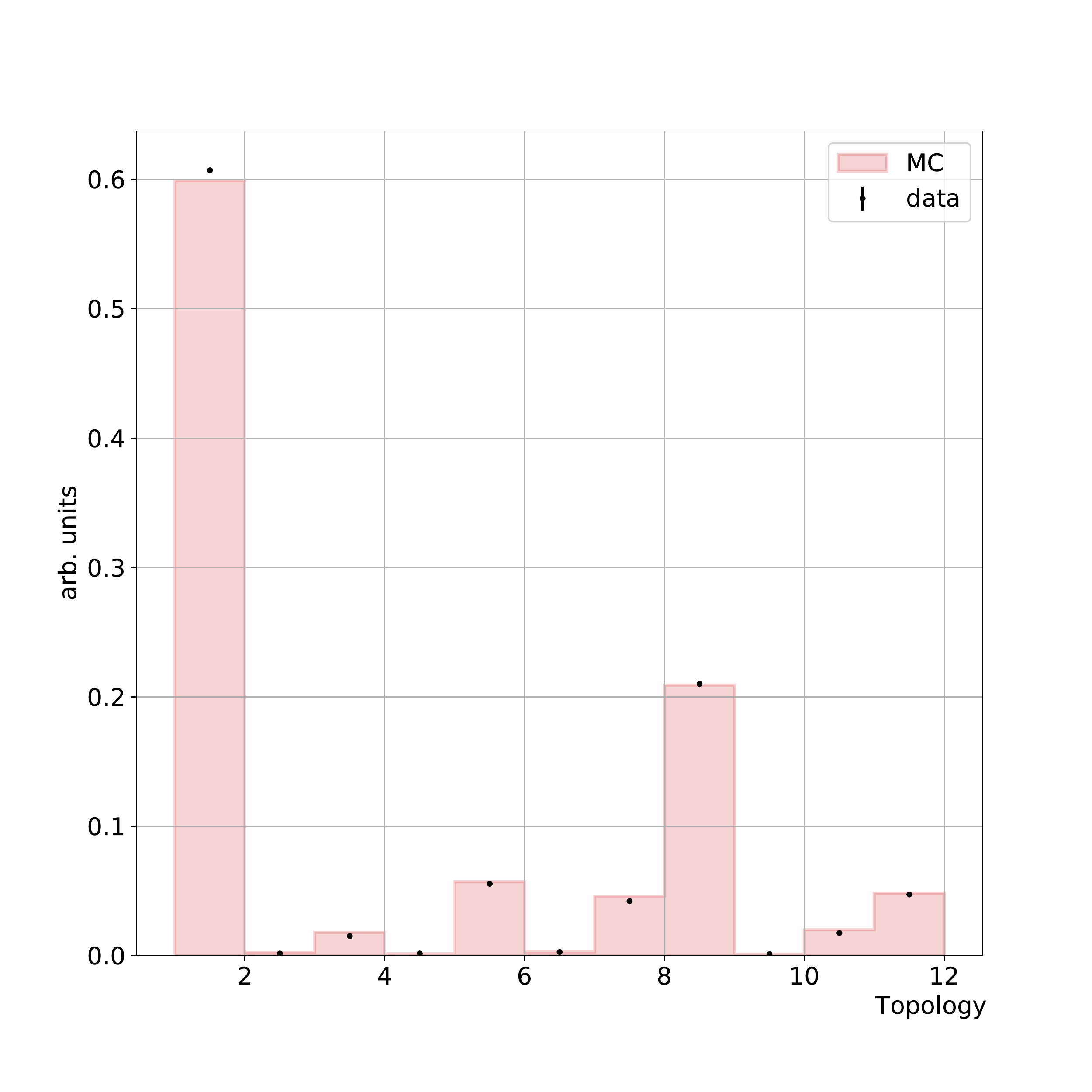}
 	\caption{\small Data to Monte-Carlo energy comparison using $^{22} \mathrm{Na}$ calibration (Left). Signal topologies recreated in the simulations (Right).}
 	\label{fig:datamc}
 \end{figure}

For the validation of the calibration and simulation work, data to Monte-Carlo comparison has been performed. Comparison using $^{22}\mathrm{Na}$ showed an agreement within 5$\%$ in the [0.2,1.2] MeV region, as shown on the Fig.~\ref{fig:datamc} (Left). Different types of events are classified into IBD topologies depending on the number and the positions of the reconstructed cubes. Those topologies are tested with $^{214}\mathrm{Bi}$ induced internal background used as a proxy for IBD. The topologies and energies induced by this background were also replicated at the percent level in simulations, see Fig.~\ref{fig:datamc} (Right).

\section{Future prospects}\label{sec:prospects}
An upgrade campaign of the detector has started in the beginning of summer 2020 with a replacement of the MPPCs to a new generation. First measurements show an increase of light yield around 40$\%$ at similar operation conditions. This upgrade is also expected to improve tagging of the IBD annihilation gammas, increase the neutron detection efficiency and
enhance background discrimination. A commissioning campaign is ongoing since the beginning of September and physics data taking will resume in the beginning of November for at least one year data taking.

\section{Conclusion}\label{sec:conclusion}
SoLid is a neutrino detector designed with a novel technology to study anti-neutrino anomalies from nuclear reactors. This technology combines hybrid scintillators, PVT and $^6$LiF:ZnS(Ag), and high segmentation. The latest is used for neutrino detection via the use of signal topologies for background reduction. An extensive calibration delivered necessary understanding of the response of the detector: its uniformity and linearity in energy response with variations of a few percent in the detector and an averaged light yield of 96 PA/MeV. After an upgrade and commissioning phase, the detector will resume the neutrino data taking for at least one more year.

\bibliographystyle{JHEP}
\bibliography{bibliography}

\end{document}